\def\lae{\mathrel{<\kern-1.0em\lower0.9ex\hbox{$\sim$}}}
\def\gae{\mathrel{>\kern-1.0em\lower0.9ex\hbox{$\sim$}}}
\def\etal{$et~al.~$}

\documentclass[12pt,preprint2]{aastex}
\usepackage{emulateapj5}
\usepackage{onecolfloat5}
\usepackage{graphicx}

\slugcomment{Accepted for publication in AJ}

\shorttitle{Intergalactic Globular Clusters in Abell 1185}
\shortauthors{JORD\'AN ET AL.}

\begin{document}
\twocolumn[
\title{A Point Source Excess in Abell 1185: Intergalactic Globular Clusters?\altaffilmark{1,2}}

\author{Andr\'es Jord\'an\altaffilmark{3}}
\affil{Department of Physics and Astronomy, Rutgers University, Piscataway, NJ 08854 \\ 
andresj@physics.rutgers.edu}
\medskip

\author{Michael J. West}
\affil{Department of Physics \& Astronomy, University of Hawaii, Hilo, HI 96720 \\ 
west@astro.uhh.hawaii.edu}
\medskip

\author{Patrick C\^ot\'e}
\affil{Department of Physics and Astronomy, Rutgers University, Piscataway, NJ 08854 \\ 
pcote@physics.rutgers.edu}
\smallskip

\and

\author{Ronald O. Marzke}
\affil{Department of Physics \& Astronomy, San Francisco State University, 
San Francisco, CA 94132 \\ marzke@quark.sfsu.edu}

\begin{abstract}
Deep imaging with WFPC2 and the {\it Hubble Space Telescope} has been used to 
search for a population of intergalactic globular clusters (GCs) belonging to 
Abell 1185, a richness class I cluster at $cz = 9800$ km~s$^{-1}$.
The field is noteworthy in that contains no bright galaxies and yet is 
centered on the peak of the cluster's X-ray surface brightness distribution. We detect
a total of 99 point sources in this field to a limiting magnitude of 
$I_{F814W} \simeq 26$. An identical analysis of the Hubble Deep Field North, which
serves as our control field, reveals a total of 12 objects in the same magnitude range.
We discuss possible explanations for this discrepancy, and conclude that 
the point source excess is likely due to the presence of GCs within A1185.
The number and spatial distribution of these GCs are consistent with their being 
intergalactic in nature, although we cannot rule out the possibility that some of the
GCs may be associated with neighboring bright galaxies. Deeper imaging with
the {\it Advanced Camera for Surveys} may resolve this ambiguity.
\end{abstract}

\keywords{Galaxies: clusters: individual (Abell 1185) -- galaxies: evolution -- 
galaxies: star clusters}

]

\altaffiltext{1}{Based on observations with the NASA/ESA {\it Hubble Space Telescope}
obtained at the Space Telescope Science Institute, which is operated by the Association
of Universities for Research in Astronomy, Inc., under NASA contract NAS 5-26555}
\altaffiltext{2}{Based in part on data obtained at the W. M. Keck Observatory, 
which is operated as a scientific partnership among the California Institute of Technology, the
University of California, and NASA, and was made possible by the generous financial 
support of the W. M. Keck Foundation.}
\altaffiltext{3}{Claudio Anguita Fellow}

\section{Introduction}
\label{sec:intro}

Although most of the stars residing in galaxy clusters are contained within distinct
galaxies, it has been known since the study of Zwicky (1951) that clusters
often have diffuse stellar components that are intergalactic in nature. The
existence of this low surface brightness starlight has been firmly established 
by numerous subsequent studies ($e.g.,$ Thuan \& Kormendy 1977; Melnick, White \& 
Hoessel 1977; V\'{\i}lchez-G\'omez, Pell\'o \& Sanahuja 1994; Bernstein \etal 1995)
but, due to the difficulties inherent in the analysis of extended, low
surface brightness features, there remains a surprising level of disagreement 
on its basic properties including spatial distribution, luminosity and color
(see the excellent review in Durrell \etal 2002).

Our ability to characterize the properties of this intergalactic stellar component 
has improved dramatically in recent years, thanks in large part to advances in instrumentation
and to  the targetted analysis of tracer
populations such as individual red giant branch stars (Ferguson, Tanvir \& von Hippel 1998;
Durrell \etal 2002), planetary nebulae (Theuns \& Warren 1997; M\'endez \etal 1997
Feldmeier, Ciardullo \& Jacoby 1998; Feldmeier \etal 2002; Okamura \etal 2002), 
type Ia supernovae (Gal-Yam \etal 2002) and even HII regions (Gerhard \etal 2002).
These studies typically find the intergalactic component to comprise 10 -- 20\% of 
the total cluster luminosity.

Globular clusters (GCs) represent an additional, and potentially powerful, tracer 
of this intergalactic component. Indeed, the possibility that galaxy clusters
may contain a population of intergalactic GCs (IGCs) has been considered many times
(van den Bergh 1958; Muzzio, Mart\'{\i}nez \& Rabolli 1984; White 1987; West \etal 1995). 
Such an IGC population
might either form {\it in situ}, as suggested by West (1993) and Cen (2001), or arise through 
the disruption or tidal stripping of cluster galaxies. Ample observational
evidence shows that these latter processes must play a role in the evolution of
galaxy clusters ($e.g.,$ Weil, Bland-Hawthorn, \& Malin 1997; Gregg \& West 1998),
as predicted by numerical simulations of cluster evolution 
(see, $e.g.$, Dubinski 1999 and references therein). 

If, as suggested by the above studies, 10 -- 20\% of the total stellar luminosity in clusters takes
the form of a diffuse intergalactic component, then it
is possible to estimate the total number of IGCs expected within a given cluster. 
Early-type galaxies 
in rich cluster environments have GC specific frequencies
(Harris \& van den Bergh 1981) of $S_N \simeq 4$ (Harris 1991).  A cluster such 
as Virgo, with a total luminosity of $L_V \simeq 1.5\times10^{12}$~$L_{V,{\odot}}$ 
inside a radius of $6^{\circ}$ (Sandage, Binggeli \& Tammann 1985)\footnote{For an
adopted Virgo distance of 16 Mpc (Ferrarese \etal 1996)} should then contain
$\simeq 10^4$ GCs associated with its intergalactic component.

However, the unambiguous detection of IGCs has proven difficult, with 
conflicting claims on whether or not such IGCs might already have been detected in nearby
clusters ($e.g.,$ West \etal 1995; Harris, Harris \& McLaughlin 1998; 
C\^ot\'e \etal 2001; Hilker 2001; Mar\' \i n-Franch \& Aparicio 2002).
Much of the confusion in this debate stems from the fact that, if IGCs follow
the same distribution within a given cluster as the other baryonic components 
($i.e.,$ cluster galaxies and X-ray emitting gas), they should be most apparent at, 
or near, the dynamical center of the cluster. Since most galaxy clusters
have a giant elliptical galaxy in close projection to their center, the
direct detection of IGCs has been hampered by the difficulties involved in 
disentangling the hypothesized population of intergalactic GCs from those 
intrinsic to the central galaxy.

The rich galaxy cluster Abell 1185 is peculiar in this respect as its brightest cluster 
galaxy is offset by $\sim$ 3$^{\prime}$, or about 120 kpc, from the centroid of its 
X-ray gas distribution (Mahdavi \etal 1996). Assuming that the X-ray centroid marks the dynamical
center of the cluster mass distribution, and is not the result of an ongoing merger,
then the detection of IGCs in Abell 1185 promises to be more straightforward than in
most clusters.

In this paper, we present deep {\it HST} observations of a field centered on the peak of 
the X-ray distribution of A1185. These observations were designed to test the IGC hypothesis
by searching for the expected point source excess compared to a blank field. As shown below,
the observations,
which were designed to sample the brightest part of the GC luminosity function at the distance 
of A1185, do indeed contain a point source excess that starts at the expected magnitude for 
an old GC population.

\section{Observations and Analysis}

We used WFPC2 on \textit{HST} to obtain F814W ($I$) imaging of A1185 as part of 
program GO-8164. The total exposure time was 13000 s, divided equally among ten, 
dithered images.  The raw data were processed with the standard STScI pipeline using the best 
available calibration files.  As a comparison field, we used a subset of the Hubble Deep 
Field North (HDF-N) observations (Williams \etal 1996), chosen to have a total exposure 
time identical to that of the A1185 field. Given the enhanced noise and smaller field 
of view of the PC chip compared to the WF chips, we limit our analysis to the latter.

Cosmic ray masks and the shifts required to register the images were obtained using the
DITHER package in IRAF.\footnote{IRAF is distributed by the National 
Optical Astronomy Observatories, which are operated by the Association of Universities 
for Research in Astronomy, Inc., under cooperative agreement with the National 
Science Foundation.} The images were combined by applying integer shifts
and averaging to produce the final images.
SExtractor (Bertin \& Arnouts 1996) was used to detect and classify 
the sources present on the field, setting the detection
threshold to 5 connected pixels having counts $2 \sigma$ above the local 
background. The FWHM of point sources, required to properly classify the
detections, was chosen on the basis of artificial star tests (see below) to be 0\farcs16.

To carry out profile-fitting photometry, and to aid in setting the parameters 
for SExtractor, we constructed point spread functions (PSFs) for our frames 
as follows. 
We used DAOPHOT (Stetson 1987) to create artificial WFPC2 frames containing 
121 stars positioned in a 11$\times$11 grid on the image.
PSFs for the F814W filter were kindly provided by P.B. Stetson.
Each artificial frame was then shifted, creating a series of images that had
the same dither pattern as the actual images for both A1185 and HDF-N. Each set 
of images was combined into a final frame using the same procedures that
were used for the actual observations. The final images created in this way
thus contain point sources constructed from the single-frame PSF but have been
subjected to the same DITHER process as the actual data.
DAOPHOT was used to determine PSFs which vary quadratically with position in the frames 
for both the A1185 and HDF-N frames.

SExtractor assigns each detected object a classification parameter,
CLASS, which ranges from
$\simeq$ 0 for galaxies to $\simeq$ 1 for point sources. CLASS is obtained
using a neural network trained on simulated images (for details see Bertin \& Arnouts 1996). 
We consider
point source candidates to have CLASS $> 0.8$. An optimized value 
for the FWHM parameter was determined by carrying out artificial star tests on the 
A1185 and HDF-N images, 
adding 25 stars at a time (so as not to alter the crowding conditions of the frame)
until a total of $5000$ stars were added to each chip. The FWHM 
quoted above was found to be the minimum value that ensures that the bulk of the artificial
stars brighter than our detection limit are classified correctly as point sources. 

\begin{figure}
\plotone{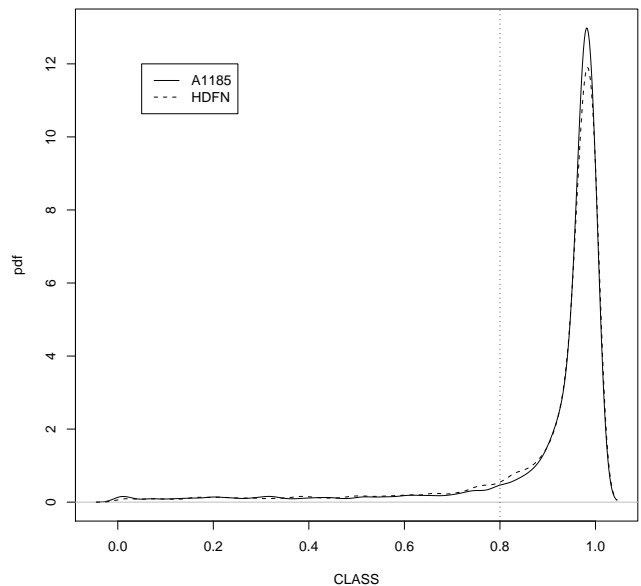}
\caption{Probability density function (PDF) for the SExtractor CLASS parameter,
recovered from the artificial star tests. The solid line shows the 
PDF for A1185 while the dashed line indicates that for HDF-N. Both distributions 
were obtained with a normal kernel density estimate. The dotted vertical line 
shows our adopted threshold for point source detections.}
\label{fig:cdist}
\end{figure}

Figure \ref{fig:cdist} shows the distribution of CLASS parameters for 
artificial stars with $I_{F814W} \leq 26$ mag recovered from the A1185 and HDF-N fields.
The majority (85\%) of these objects have CLASS $\geq 0.8$. 
The fraction of misclassified artificial
stars depends sensitively on the magnitude cutoff, falling to $\sim$ 8\% for 
$I_{F814W} \le 25.5$ and $\sim$ 5\% for $I_{F814W} \le 24.5$. Based on these 
artificial star tests, we find the $50\%$ completeness level lies at 
$I_{F814W} \sim 26$ mag for A1185 and $I_{F814W} \sim 26.2$ mag for HDF-N.
As Figure~\ref{fig:cdist} makes clear, the distributions for the two fields
are similar, as is necessary if we are to compare directly the counts for the 
two fields.

PSF-fitting photometry was then carried out with DAOPHOT, taking
the sky from an annulus with inner and outer radii of 5 and 10 pixels.
Instrumental magnitudes were transformed into the VEGAMAG system using zero-points
taken from the \textit{HST} WFPC2 Data Handbook (Baggett \etal 2002). 
A correction for foreground extinction was performed using the reddening 
curves of Cardelli, Clayton \& Mathis (1989), with the value of $E(B-V)$ taken 
from the DIRBE maps of Schlegel, Finkbeiner \& Davis (1998). A final correction
of $0.1$ mag was applied to correct from a PSF radius of $0\farcs5$ to one of 
infinite aperture (Holtzman \etal 1995). 

Those objects having a DAOPHOT $\chi$ statistic\footnote{$\chi$ is a robust estimate of the 
ratio of the observed to expected scatter about the model profile (Stetson \& Harris 1988).}
greater than $1.75$ were discarded and 
any point sources brighter than $3\sigma$ above the expected turnover magnitude 
of the globular cluster luminosity function at the distance of A1185 were
omitted from the analysis. The threshold on $\chi$ was determined 
on the basis of the artificial star tests.
We assumed $M_{I,TO} = -8.4$ mag and $\sigma = 1.4$ mag for 
the absolute magnitude of the turnover and the dispersion of the Gaussian 
representation of the GC luminosity function, respectively (Harris 2002). The 
distance to A1185 is calculated directly from its redshift, $z=0.0325$, assuming 
$H_0=72$ km s$^{-1}$ Mpc$^{-1}$ (Freedman \etal 2001). This procedure results in
the rejection of all objects brighter than $I_{F814W} = 23.1$ mag and eliminates
$9$ objects in the A1185 field and $10$ in HDF-N. For comparison, $6-7$ galactic stars
with $I \lesssim 23 $ are predicted by the Bahcall-Soneira  
Galaxy model (Bahcall \& Soneira 1980; Bahcall 1986) in these fields

As a final check on the point-source nature of the remaining detections, 
we examined the radial profile of the difference between the observed profile
of each source with the PSF profile expected at that position in the field. 
While a trend
between the median difference and radius might then be expected if the sources 
were extended no such trend was detected, supporting the SExtractor 
classifications.

\section{The Nature of the Excess}
\label{sec:res}

\begin{figure}
\plotone{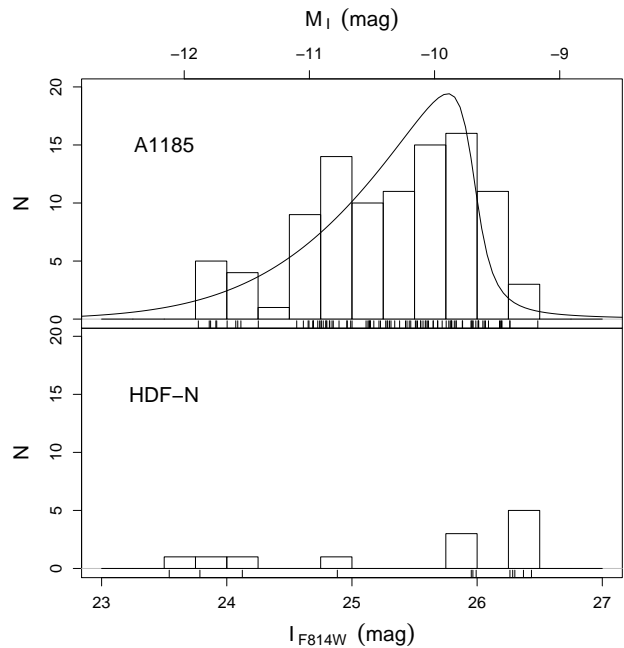}
\caption{{\it (Top Panel)} Luminosity function of point sources detected in A1185. 
The solid line is the expected luminosity function of GCs at the distance of 
A1185, multiplied by the completeness function and normalized to the observed counts. 
{\it (Bottom Panel)} Luminosity function of point sources detected in the HDF-N field.}
\label{fig:hists}
\end{figure}

Our final catalog consists of 99 point sources in the A1185 field. By
contrast, we find 12 sources in the HDF-N field, which serves as our control. 
The luminosity functions for both fields are shown in Figure~\ref{fig:hists}. For A1185
we also show the expected luminosity function for GCs at the distance of A1185,
multiplied by the completeness function and normalized to the observed number of sources.
Our counts for the HDF-N field are in good agreement with those of Elson, 
Santiago \& Gilmore (1996) who find 14 point sources in the same magnitude interval.
In what follows, we examine a number of possible explanations for this point source
excess.

\subsection{Galactic Field Stars}

We first consider the possibility that the A1185 field
contains a larger number Galactic field stars than does the HDF-N field. However, 
this explanation appears 
unlikely given the size of the excess ($i.e.,$ nearly an order of magnitude) and 
the roughly similar Galactic latitudes of the two fields: $b$ = 67\fdg8 for A1185 and 
$b$ = 54\fdg8 for the HDF-N field. Indeed, according to the Bahcall-Soneira
Galaxy model (Bahcall \& Soneira 1980; Bahcall 1986), both fields are predicted to 
contain only 4-6 stars in the magnitude range $23 \lesssim I \lesssim 26$. 
We conclude that Galactic field 
stars are unlikely to be the origin of the observed excess.

\subsection{Background Galaxy Cluster}

Could the excess be due to a distant background galaxy cluster which
happens to fall in the A1185 field? The low surface density of
high-redshift clusters (e.g. Postman \etal 2002)
suggests that this explanation is unlikely. More importantly, we expect
almost all background giant ellipticals to be resolved on our WF
frames:  i.e., the angular diameter distance turns around at $z \sim 1.5$ in
such a way that, for $z \gtrsim 1$, 1$^{\prime\prime}$
corresponds to $\simeq$~6~kpc. For instance, Lyman break galaxies at
$z \sim 3$, though physically compact, are still clearly resolved in
deep WFPC2 images, with half-light radii $\gae$ 0\farcs2 (Steidel \etal
1996). A definitive test should be possible with
multi-color data (and particularly IR colors) for the point source
population in A1185.

\subsection{Nuclei of dE,N Galaxies}

A third possibility is that our point source catalog includes a population 
of dE,N galaxies. At the distance of A1185, the bright nuclei of such galaxies 
would be unresolved.
As Binggeli \& Cameron (1991) have shown, the nuclei of dE,N galaxies mimic the
bright end of the GC luminosity function in the Virgo cluster, so it is conceivable 
that our sample may include some dE,N galaxies, particularly since our WFPC2 
field is located near the center of A1185 where the galaxy surface density 
is expected to be high.
There are two separate issues that must be addressed to test the plausibility of 
this explanation: (1) the probability that dE,N galaxies would be classified
incorrectly by SExtractor as point sources; and (2) overall number of dE,N galaxies
expected in our A1185 field. This latter point will be examined in detail below.

To address the first issue, we carried out simulations in which artificial 
dE,N galaxies were added to our frames. These frames were then run through the same 
reduction procedures as the actual observations and, at the end of each run, 
the classifications returned by SExtractor were recorded. The dE,N galaxies were simulated 
using the data presented in Table 1 of Lotz \etal (2001) which gives total magnitudes, 
exponential length scales, nuclei $V$-band magnitudes and $V-I$ colors for a sample
of 27 dE,N galaxies in the Virgo and Fornax clusters that have been observed with
\textit{HST}. To simulate the dE,N galaxies, we first used DAOPHOT to add point 
sources with the tabulated $I$ magnitudes after shifting to the distance of A1185.
The task MKOBJECTS in IRAF was then used to add an exponential disk having the 
appropriate exponential scale, ellipticity and  total magnitude (minus the contribution of the 
nucleus). The colors of the exponential disks were taken to be $(V-I) = 0.65 - 0.022M_V$ and 
$(B-V) = 0.41 - 0.022M_V$; these relations were obtained by combining a 
[Fe/H]--$M_V$ relation for dwarf galaxies (C\^ot\'e \etal 2000) with the 
color--metallicity relations presented in Barmby \etal (2000).
Each galaxy in the Lotz \etal (2001) sample was simulated
$50$ times per WF chip, with the galaxies being simulated in groups of $9$ so 
as not to overcrowd the fields. These simulations revealed that all dE,N
galaxies were classified correctly as non-stellar in all trials. We therefore conclude
that the nuclei of dE,N galaxies cannot be the origin of the point source excess
in A1185.

\subsection{GCs associated with Dwarf Galaxies}

A related possibility is that the point sources detected in the A1185 field represent 
bona fide GCs, but ones that are associated with dwarf galaxies that may be present in 
A1185. We examine this possibility by calculating the expected number of dE and dE,N 
galaxies in the A1185 field.  Since the velocity dispersions of Virgo and A1185
are similar (Binggeli, Tammann \& Sandage 1987; Mahdavi \etal 1996), their dwarf
populations can be compared directly. Based on the Virgo dE+dE,N luminosity
function published by Sandage, Binggeli \& Tammann (1985), 
expect $\sim 1000$ dE and dE,N galaxies in Virgo out to a distance of $5^{\circ}$. We 
have adopted a cutoff at the faint end of the luminosity function of $M_B=-11.5$ 
as there are no known GC systems belonging to dwarfs fainter than that.
Given that A1185 is $\sim 9$ times more distant than Virgo, and assuming that the surface 
density of dwarfs in Virgo has the approximate form $e^{-r/\alpha}$ with
$\alpha \simeq 1\fdg7$ (Ferguson \& Sandage 1989) \footnote{This value of $\alpha$ 
is appropriate for dE + dE,N fainter than
$M_B \sim -13$. Brighter dwarfs exhibit in general larger $\alpha$ and this would drive 
down the estimated number of dwarfs in our field.}, 
the expected dE surface density at the center of A1185 is roughly 5500 deg$^{-2}$. 
Thus, we expect
our WFPC2 field to contain $\lesssim 8$ dE and dE,N galaxies. Given that these galaxies 
would typically contain only $\simeq 5$ clusters each 
($e.g.,$ Lotz \etal 2001), and that our 
photometry samples only the brightest 16\% of the GC luminosity function
at the distance of A1185, we expect only  $\lesssim$ 8 point sources to be GCs associated 
with dwarf galaxies in A1185.

\subsection{Faint dE galaxies}

We have also explored the possibility that the point source catalog in A1185
has been contaminated by faint dE galaxies with small exponential scale-lengths. 
This possibility has been
investigated by adding simulated dE galaxies to the WFPC2 frames, with the
galaxies modeled as pure exponential disks having scale-lengths of 
$r_e = 0.1$~kpc. This choice of scale-length corresponds to the low-end of
the values found for Local Group dwarfs (Mateo 1998). It is these highly
concentrated dwarfs that are cause for the greatest concern, as they are
the ones most likely to be misclassified at the distance of A1185. We added
150 artificial dE galaxies having absolute magnitudes 
$M_I = -11.7,~-11.2,~10.7,~-10.2,~-9.7$ and $-9.2$ (1200 in total) to each 
WF chip and analyzed the images in the same way as the actual data. Only
for $M_I = -10.7$ and $-10.2$ were any dwarfs misclassified, and then only
in $\sim$1\% of the cases. Combining this result with the fact that
we expect $\sim$ 8 dE galaxies in this magnitude range (determined from
an extrapolation of the fitted Schechter function for the dE population
in Virgo; Sandage \etal 1985), we find that the number of compact dE galaxies
in our point source catalog is completely negligible.

\subsection{A1185 Globular Clusters}

\begin{figure}
\plotone{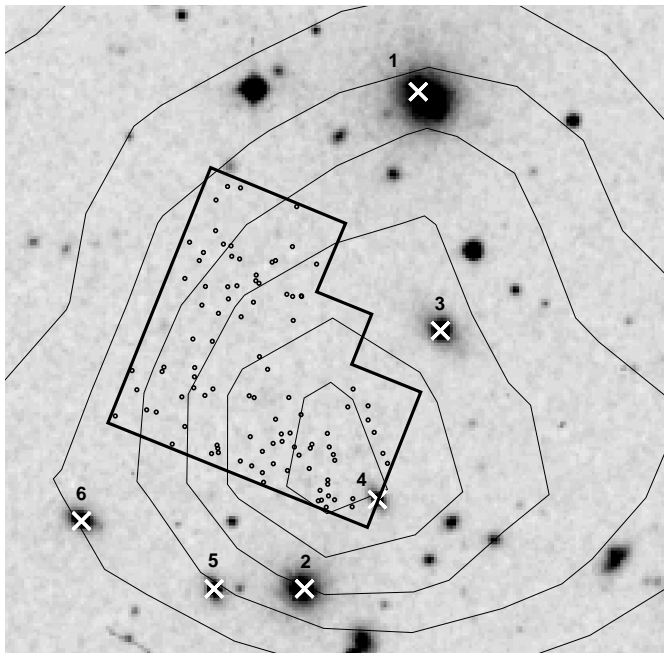}
\caption{Digitized sky survey image of A1185, showing the WFPC2 field and
detected point sources (circles). North is up and East is to the left in this 
image, which measures 6\farcm0$\times$6\farcm0. Crosses denote the six galaxies
whose globular cluster systems are modeled in \S~3. The numbering is as described
in the text. The thin lines shows 
contours of constant X-ray surface brightness based on Einstein IPC observations.}
\label{fig:fov}
\end{figure}

The calculations above suggest that the number of dwarf galaxies and Galactic field 
stars expected in the A1185 field fall far short of that needed to explain the point source 
excess. We conclude that the point source excess is probably the result of bona fide GCs 
in A1185, but is it plausible that the sources are GC associated with nearby ellipiticals
rather with the cluster as a whole? As Figure~\ref{fig:fov} shows, a few of the sources
are close in projection to neighboring galaxies, but the bulk of the population is not 
obviously associated with galaxies. To examine this issue more quantitatively, we 
consider the GC systems of the six bright galaxies that are nearest to the WFPC2 
field: NGC~3550 (BCG), NGC~3552, CGCG 155-081, 2MASSXi J1110398+284224,
MCG+05-27-004 NED01 and MCG+05-27-004 NED02. These six objects are marked by crosses
in Figure~\ref{fig:fov}, and labelled according to the above order.

To estimate the number of GCs that are associated with these galaxies and fall in the
WFPC2 field, it is first necessary to 
assume a radial density profile for their GC systems. On theoretical grounds, the 
profiles of non-central galaxies are expected to be tidally truncated by the mean 
cluster field ($e.g.,$ Merritt 1984), but precise values for the truncation radii
of cluster galaxies have not been determined observationally, nor has the process been
probed adequately in numerical simulations. As a first estimate, we use the theoretical predictions 
for the tidal radius put forward in Merritt (1984), assuming $\alpha=\beta=1.5$ in his 
equation (6a). The cluster velocity dispersion is taken to be $\sigma_v=736$ km~s$^{-1}$ (Mahdavi 
\etal 1996) and the velocity dispersion for each galaxy is estimated using the Faber-Jackson 
relation (Faber \& Jackson 1976). For the radial density profiles inside the tidal 
radius, we use power laws of the form $r^{\alpha}$ with $\alpha$ given by 
\begin{equation} \alpha = -0.29M_V - 8.0 \label{eq:alfa} \end{equation}
(Kaisler \etal 1996). 
Absolute magnitudes, $M_V$, were measured for the A1185 galaxies by running SExtractor 
on a $30$s $V$-band exposure taken with the Low-Resolution Imaging Spectrometer 
(Oke \etal 1995) 
at the W.M Keck Observatory on 19 March, 1999.
To estimate the total number of GCs belonging to each galaxy, we assume 
a constant GC specific frequency of $S_N = 4$ (Harris 1991). Integrating 
the various GC density profiles over the WF chips, and taking into account 
the fact that we observe only the brightest $\sim 16\%$ 
of the GC luminosity function, we find a total contribution 
of $\sim 15$ GCs from the GC systems of these galaxies.\footnote{
The galaxy 2MASSXi J1110398+284224 lies partly on the WF4, 
so we neglect any GCs falling within 10$^{\prime\prime}$ of its center
since these GCs would go undetected in our reduction pipeline.}
Of the 15 expected GCs, $13$ would be on WF4. Although this detector is indeed 
observed to contain largest number of point sources (39), the factor of six 
discrepancy between the observed and predicted counts suggest that the point
source excess is unlikely to be due to GCs associated with neighboring galaxies,
if the galaxies are tidally truncated as described above.

However, this estimate depends rather sensitively on the assumed recipe for 
tidal limitation. We have therefore calculated the expected contribution 
when this effect is neglected entirely. Since we are now interested in the 
spatial distribution of GCs at very large projected radii, the power-law 
representations of the radial density is inadequate (Rhode \& Zepf 2000). 
We instead assume that all galaxies have GC surface density profiles 
similar to the $r^{1/4}$ profile of M49. Beyond
$r = 100$ kpc, the profiles are set to zero, consistent with the results of 
Rhode \& Zepf (2000). This is a rather conservative approach, as the fainter galaxies
will certainly have less extended profiles (see Eq.~\ref{eq:alfa}). Following the same 
procedure as described above, we 
then find an expected contribution of $\sim 62$ GCs, still lower than the observed number 
of point sources, but now within a factor of $\sim$ 50\% of the actual excess. We 
caution, however, that 
the previous estimate is probably more realistic, as the cluster galaxies have almost
certainly been tidally truncated at some level. Indeed, dynamical modeling of the
GC system of M87 --- the central elliptical in the Virgo cluster --- has shown the 
velocity dispersion profile of its GC system to rise beyond $r \sim 20$ kpc, as expected if the
GCs are orbiting in the potential well defined by both the galaxy and its parent cluster
(C\^ot\'e \etal 2001). At a projected distance of 100 kpc, the gravitational potential in which
the GCs orbit is dominated not by the central galaxy, but by the cluster itself.

Could a population of IGCs be responsible for the observed point source excess?
West \etal (1995) proposed a phenomenological model in which the high specific frequencies
of some centrally dominant galaxies ($e.g.,$ Harris 1991) are explained by the presence of
a population of IGCs. In this scenario, putative populations of IGCs provide an ``excess"
GC population that elevates the observed specific frequencies above a universal value 
of $S_N = 4$.  Following West \etal (1995) and Blakeslee \etal
(1997), we assume that the surface density of IGCs is given by
$$\Sigma_{\rm IGC} = {\Sigma_{0}}T_X/(1+r^2/r_c^2)$$ where
where $r_c$ and $T_X$ are the cluster's core radius and X-ray temperature, respectively.
To find the constant of proportionality in this relation, we performed a linear fit
to the data presented in Figure~13 of Blakeslee, Tonry \& Metzger (1997).
Taking into account that the area of their detector was 22 arcmin$^2$, the
predicted IGC surface density is
$$\Sigma_{\rm IGC} = 33 T_X/(1+r^2/r_c^2)~{\rm arcmin^{-2}}$$
Using a value of $T_X = 3.9$ keV for A1185 (Jones \& Forman 1999) and
recalling that we are sensitive to only the brightest $\sim 16\%$ of the GC
luminosity function, we expect our WFPC2 field to contain  $\sim 110$ IGCs.
The predicted number of IGCs is thus of the right order needed to explain
the point source excess in A1185.

\section{Conclusions}
\label{sec:dis}

Having considered the expected contributions of many different sources, it seems clear
that the observed point source excess is best explained by a population of GCs in A1185.
Whether these GCs are intergalactic in nature, or whether they are associated with 
neighboring galaxies, is an open question. 
The difficulty in deciding between these scenarios lies in the unknown extent to which 
the cluster mean field has imposed a tidal limitation on the individual cluster galaxies.
If tidal truncation has been less effective than assumed in the above analysis, then 
many, and perhaps most, of the observed point sources observed might be explained by GCs 
associated with individual galaxies. On the other hand, if the cluster galaxies
have been tidally limited as prescribed in Merritt (1984), the IGC hypothesis emerges as 
the most plausible explanation for the observed excess. 

Based on the evidence above, it seems very likely that the point sources observed in 
our WFCP2 field are bonafide GCs residing in A1185. Nevertheless, this assertion can be tested
with deeper imaging of this field. Any population of GCs, whether or not it is intergalactic in
nature, should be characterized a near-Gaussian luminosity function. At the distance
of A1185, the GC luminosity function is expected to show a turnover at $I_{TO} \simeq 27$,
which is well within the imaging capabilities of {\it HST} and the Advanced Camera for 
Surveys. While the luminosity function alone would would not allow us to distinguish between
GCs belonging to individuals galaxies and those associated with the cluster
as a whole, it may be possible to discriminate between these alternatives by examining 
the spatial distribution of an expanded sample of sources.
Additional clues to the nature of these GCs, including ages and metallicities,
will be possible once optical and infrared colors are in hand.

\acknowledgments

We would like to thank P.B. Stetson for providing PSFs for the F814W filter and 
an anonymous referee for helpful comments.
Support for program GO8164 was provided by NASA through a grant from the Space Telescope 
Science Institute, which is operated by the Association of Universities for Research in 
Astronomy, Inc., under NASA contract NAS5-26555.  Additional support for this work was 
provided by the National Science Foundation from grant AST-0205960 and through a grant from the
Association of Universities for Research in Astronomy, Inc., under NSF cooperative
agreement AST-9613615 and by Fundaci\'on Andes under project No.C-13442.
This research has made use of the NASA/IPAC Extragalactic Database (NED) which is 
operated by the Jet Propulsion Laboratory, California Institute of Technology, under
contract with the National Aeronautics and Space Administration. This work was based in 
part on observations obtained at the W. M. Keck Observatory, which is operated jointly by the
California Institute of Technology and the University of California. We are grateful to 
the W. M. Keck Foundation for their vision and generosity.

%
%

\end{document}